\documentclass[useAMS]{mn2e}
\usepackage{psfig}
\usepackage{graphicx}

\title[Superdisks in Radio Galaxies]{Superdisks in Radio Galaxies: Jet--Wind Interactions }

\author[Gopal-Krishna et al. ] 
{Gopal-Krishna$^{1}$\thanks{E-mail: krishna@ncra.tifr.res.in (G-K),
wiita@chara.gsu.edu (PJW), santosh@iucaa.ernet.in (SJ)},  Paul J.\ Wiita$^{2}$\footnotemark[1], 
Santosh Joshi${^3}$\footnotemark[1]\thanks{On leave from ARIES, Nainital, 263129, India}
  \\  
$^{1}$ National Centre for Radio Astrophysics, TIFR,  Post Bag 3, Pune 411
007, India\\
$^{2}$ Department of Physics \& Astronomy, Georgia State University, P.O. Box 4106, Atlanta,
Georgia 30302-4106, USA \\
$^{3}$ Inter-University Centre for Astronomy and Astrophysics (IUCAA), Post Bag 4, Ganeshkhind, Pune 411 007, India}

\date{Accepted 2007 June 12.
      Received 2007 June 8; in original form 2006 May 24}

\pagerange{\pageref{firstpage}--\pageref{lastpage}}
\pubyear{xxx}

\begin{document}

\maketitle

\label{firstpage}

\begin{abstract}
Taking a clue from their sharp-edged (strip-like) morphology observed in 
several cases, a new mechanism is proposed for the formation of the
emission gaps seen between the radio lobes of many powerful extragalactic 
double radio sources.  
Canonical understanding of the radio gaps invokes either blocking of the 
back-flowing lobe plasma by the denser interstellar medium (ISM) of the host 
galaxy, or ``squeezing" of the radio bridge in the middle through buoyancy 
force exerted by either the ISM or the surrounding intra-cluster medium (ICM).
These pictures encounter difficulties in explaining situations where  the 
sharp-edged radio gaps associated with non-cluster radio galaxies have widths 
running into several tens (even hundreds) of kiloparsecs. More particularly, 
the required dense high-pressure ISM/ICM is likely to be lacking at least
in the case of high-redshift radio galaxies. We propose here that radio 
emission gaps in at least such cases could arise from a dynamical 
interaction between the powerful thermal 
wind outflowing from the active galactic nucleus and the back-flowing 
synchrotron plasma in the two radio lobes, which occurs once the rapidly 
advancing jets have crossed out of the wind zone into the intergalactic 
medium. A simple analytical scheme is presented to explore the plausibility 
of the side-ways confinement of the thermal wind by the radio lobe pair, 
which would ``freeze" pancake shaped conduits in the space, along which the 
hot, metal enriched wind from the AGN can escape (roughly orthogonal to the 
radio axis). Some other possible consequences of this scenario are pointed out.
\end{abstract}


\section{Introduction}
Over the past few decades, aperture-synthesis observations of radio 
galaxies have yielded, with increasing precision, maps 
of the synchrotron plasma deposited by the jet pair near their extremities 
(hot spots) as they advance through the ambient medium. Already from 
comparison of the early high-resolution observations made at centimetre 
and metre wavelengths (Jenkins \& Scheuer 1976; Gopal-Krishna \& Swarup 
1977),
it was strongly hinted that the the pair of radio lobes arising from the 
backflow from 
the two hotspots are often separated by an emission gap which is probably 
real, and not merely a result of radiative losses.  Such emission gaps,
particularly the largest of them, 
will be the focus of the present
study. According to one interpretation, the central gap in the ``radio 
bridge" arises from blocking of the backflowing radio plasma by the 
denser interstellar medium (ISM) of the parent elliptical galaxy (e.g., 
Leahy \& Williams 1984; Alexander 1987; Wiita \& Norman 1992). 
However, the general applicability of this mechanism was found to be at 
odds with the observation that in several powerful radio galaxies at moderate
redshifts, the radio gaps have sharp, quasi-linear edges and some of these 
have widths approaching or even exceeding 100 kpc (Gopal-Krishna \& Wiita 1996;
Gopal-Krishna \& Nath 1997; Gopal-Krishna \& Wiita 2000, hereafter GKW00).
 
In these papers we argued that such emission gaps, or ``superdisks", can provide 
consistent explanations for some key correlations found among radio source 
properties, such as the Laing-Garrington effect (Garrington et al.\ 1988; 
Laing 1988; Garrington, Conway \& Leahy 1991), the correlated radio-optical 
asymmetries (e.g., McCarthy, van Breugel \& Kapahi 1991), 
the preference of multiple absorption dips to occur in the Ly-$\alpha$ profiles of high-$z$ radio galaxies with overall sizes below
$\sim$35 kpc (van Oijk et al.\ 1997; Binette  et al.\ 2006) and metrewave flux 
variability via ``superluminal refractive scintillations" (Gopal-Krishna 1991; 
also, Campbell-Wilson \& Hunstead 1994; Ferrara \& Perna 2001). Another 
interesting property hinted by the GKW00 sample of superdisks was that the hot 
spots in the radio lobe pair are usually located more symmetrically about the 
superdisk's mid-plane than about the host galaxy, possibly signifying the 
galaxy's motion during the AGN phase (see Sect.\ 4).  

In high redshift radio galaxies, the apparent asymmetry of diffuse 
Ly$\alpha$ emission, 
which is a sensitive tracer of dust, can be understood if the brighter 
Ly$\alpha$ emission is associated with the radio lobe on the near side of 
the nucleus and thus not obscured by any dust present in the superdisk (GKW00). 
Because the sharp-edged morphology would only be noticed when the radio jets
are quite close to the plane of the sky, they are subject to a strong negative
selection effect, and GKW00 argued that even though the observed cases of
superdisk are few, the phenomenon may not be rare.

Initially we proposed that the superdisk is primarily made of the 
interstellar medium bound to the radio galaxy (RG) itself, perhaps 
originating from the gas belonging to gas-rich disk galaxies previously 
captured by the giant elliptical host of the powerful RG (e.g., Statler 
\& McNamara 2002). We argued that the tidal stretching and heating of that
gas during the capture was sufficient to produce very large fat pancakes 
(Gopal-Krishna \& Nath 1997; GKW00). Since in some extreme cases (e.g., 
0114$-$476; Saripalli, Subrahmanyan \& Udaya Shankar 2002) the width of the superdisk was found 
to run into several hundred kiloparsecs, an alternative possibility was 
also put forward in GKW00, according to which at least some superdisks 
trace the gaseous filaments of the ``cosmic web".   Earlier, noticing the 
straight, abrupt inner boundaries of the lobes in a radio galaxy (3C~227), 
Black et al.\ (1992) had inferred a ``disc of central dense, cold gas 
with axis coincident with that of the source". Very recently, Gergely
\& Biermann (2007) have proposed that superdisks can be carved out by 
the powerful wind produced due to a rapid jet precession during the 
impending merger of two supermassive black holes belonging to a pair
of merging galaxies. 

A  popular explanation for the emission gaps, in general,
invokes a squeezing or pinching of the (lighter) synchrotron 
plasma of the radio bridge in the middle, 
by the denser, higher pressure ISM, or circum-galactic medium, associated with the parent galaxy (e.g., Scheuer 1974; 
Croston et al.\ 2004b). 
While it is presently unclear how this process can account for the
gaps with sharp, straight boundaries, the existence of such a medium is 
indeed supported by the detection of resolved X-ray emission from many massive ellipticals hosting 
double radio sources at small to medium redshifts (e.g.\ Hardcastle \& Worrall 2000). 
This ISM is likely to be built up through a gradual accumulation of the
gas lost by the stellar population over the Hubble time and heated to 
$\sim 10^6$ -- $10^7$ K due to mixing within the galaxy dominated by random 
motions (Brighenti \& Mathews 1997 and references therein; O'Sullivan, 
Forbes \& Ponman 2001). 

On the other hand, massive ellipticals at high redshifts  are expected to lack 
a significant amount of high-pressure ISM. Then the medium around even massive 
galaxies situated in groups and clusters will resemble peaks in the 
intergalactic medium (IGM) (and not the higher pressure intracluster medium 
which is typical of their lower redshift counterparts) due to an early stage 
of gas accretion from the cosmic web and its virialization.
Based on sensitive optical/X-ray observations and cosmological simulations,
clusters at $z >~ 1.5 $ are believed to be proto-clusters which would have 
grown into the present-day clusters by the process of virialization and 
merger with other clusters or galaxy groups (e.g., Miley et al.\ 2004; 
Ford et al.\ 2004; also, Rowley, Thomas \& Kay 2004; Motl et al.\ 2004; Tormen, Moscardini \& Yoshida 
2004). 
Significant concomitant enrichment of the ICM is also expected via galactic
winds and ram-pressure stripping of the galaxy population (e.g., Kapferer et
al.\ 2007). 

Our premise that distant radio galaxies have a circum-galactic medium at a 
significantly lower temperature/pressure than do nearer ones is supported by the deep {\sl Chandra} 
observations which show that spectroscopically selected high redshift ($z \sim 
1$) clusters and groups of galaxies are under-luminous, compared to their low 
redshift counterparts (Fang et al.\ 2006). A similar result is 
reported for medium redshift groups of galaxies (Spiegel, Paerels \& Scharf 2007). The lower X-ray output probably indicates that these high-$z$ 
systems have yet to accrete adequate hot gas and/or being dynamically younger 
systems, they are not yet virialized.  It is thus expected that the (forming) clusters
at high redshifts would be lacking a dense high-pressure medium found
within nearby clusters and groups of galaxies. It may be noted that although X-ray emission extended on 
at least 100 kpc 
scales has been detected around the $z= 1.786$ quasar 3C294 and
the $z=2.48$ RG 4C 23.56 from deep {\sl Chandra} observations, it is more likely
to be inverse Compton boosted cosmic microwave background, as its energy density is much greater at such high redshifts (Fabian et al.\ 2003; Johnson et al.\ 2007; see also,
Celotti \& Fabian 2004). These considerations motivate us 
to explore in this paper an alternative mechanism that could be effective 
in the formation of emission gaps between the radio lobes in higher redshift 
radio galaxies.

Non-relativistic winds, with speeds often exceeding 10$^3$ km s$^{-1}$ 
and a mass outflow rates often greater than 1 M$_{\odot}$ yr$^{-1}$ (e.g., 
Soker \& Pizzolato 2005; Brighenti \& Mathews 2006; Crenshaw \& Kraemer 2007;  
Tremonti, Moustakas \& Diamond-Stanic 2007) are now thought to be an integral part of the active 
galactic nucleus (AGN) phenomenon, and can be launched from accretion disks 
via several mechanisms (e.g., Narayan \& Yi 1994; K\"onigl \& Kartje 1994; 
Blandford \& Begelman 1999; Das et al.\ 2001).
Evidence for such outflows comes from their absorption of the underlying 
continuum in the UV and/or X-rays (e.g., PDS 456, Reeves, O'Brien \& Ward 
2003; PG 1211$+$143, Pounds et al.\ 2003; NGC 1097, Storchi-Bergmann et al.\ 
2003; see Crenshaw, Kraemer \& George 2003 for a review).  Such relatively 
slow winds and relativistic jets may coexist in many objects (e.g.\
Binney 2004; Gregg,
Becker \& de Vries 2006). 
As these non-relativistic winds, together with any existing relativistic jets,
are believed to carry a mechanical luminosity far in excess of the synchrotron
luminosity (e.g., Willott et al.\ 1999), they are capable of significantly 
influencing important phenomena such as cooling flows in clusters (e.g., Omma 
\& Binney 2004; Soker \& Pizzolato 2005; Brighenti \& Mathews 2006), and even 
the structure formation in the universe (e.g., Rawlings 2003; Gregg et al.\ 
2006). It is certainly possible that the jets contain significantly more 
mechanical energy than do the winds (Rawlings 2003), and that the mechanical
power extracted from the central engine can sometimes greatly exceed its 
bolometric luminosity (Churazov et al.\ 2002, 2005; Peterson \& Fabian 2006).
An interesting question is: {\it what sort of interactions might occur 
between these two kinds of outflows of AGN power?}  Here we examine this 
issue  and 
propose that at least some of the  radio emission gaps, particularly in 
radio galaxies at higher redshifts, could well be  manifestations of such an 
interaction.

Association of non-relativistic outflows with the accretion process has 
been discussed in many studies. An example is the ``advection dominated 
inflow-outflow solution'' (ADIOS)  model (e.g., Blandford \& Begelman 1999). 
In this model, most of the gravitational energy released during accretion
 onto a BH gets used up in driving a wind from the surface of the accretion 
disk.  Thus, most of the gas falling on the outer edges of the disc is carried
away by the wind roughly at the local Keplerian speed.  In contrast, 
relativistic jets in RGs may be  energized by the spin of the BH (e.g., 
Blandford \& Znajek 1977; also, Wilson \& Colbert 1995) and are almost 
certainly launched from the innermost portion of the central engine (e.g., 
Meier 2003). The ejection of relativistic jets probably marks a later phase in 
The accretion process (e.g., Rawlings 2003; Gregg et al. 2006).  Another key 
difference between the two kinds of outflows is that the non-relativistic 
wind is likely to be much less collimated and, on the scale of 100 kpc, it 
almost certainly could be regarded as quasi-isotropic (e.g., Levine \& 
Gnedin 2005). For the purpose of the exploratory calculations presented here 
we shall follow these last authors and adopt a spherical geometry for the 
non-relativistic wind outflow.  We then proceed to analytically compute the 
interaction of the radio jets with an older phase of AGN activity which leads 
to winds that can blow up large ``bubbles'' of hot thermal gas around the 
elliptical galaxy hosting the AGN.  Thus, in this picture, a central radio 
emission gap or even a superdisk in a high-$z$ RG is created by the RG 
itself, but it is not predominantly comprised of the host galaxy's interstellar 
medium (either captured or due to stellar winds).  In Section 2 we discuss 
the basic idea and note some relevant observations.  Section 3 presents a
simple analytical framework for this two phase model of AGN activity. Some 
possible implications of this scenario are discussed in Section 4.

\section{Radio Galaxies with Central Emission Gaps}

Assume the formation of a powerful AGN begins with the launching of a 
thermal gas outflow from the nuclear region in the form of bubbles which 
coalesce into a single bubble (e.g.\ Di Matteo, Springel \& Hernquist 2005; 
Gregg et al.\ 2006). As this bubble of hot thermal wind expands it approaches 
a roughly spherical shape; it also will eventually tend to reach pressure 
equilibrium with the IGM.  At some later time, but typically before the 
bubble/IGM equilibrium is reached, the radio jets are launched from the 
same AGN in two opposite directions. 

We  reiterate that the situation will be significantly different for 
radio galaxies forming  in situations where hot, X-ray emitting gas at 
significantly higher pressure is the norm, such as those within well 
developed clusters or groups of galaxies, or in massive ellipticals at low 
redshift.  Under those circumstances much smaller bubbles of a few kpc 
radius will normally be inflated by the winds and/or radio jets (e.g.\ Dunn \& 
Fabian 2006; Sternberg, Pizzolato \& Soker 2007).  Although these bubbles 
may not lead to substantial density changes in the circumgalactic medium, 
they would give rise to patchiness in the X-ray emission. Examples of such
patchiness can be found in the {\sl Chandra} images of some nearby X-ray 
bright clusters, such as Perseus (Fabian et al.\ 2000), A262 (Clarke et al.\ 
2007) and 2A0335$+$096 (Mazzotta, Edge \& Markevitch 2003). These clusters 
provide examples where X-ray patchiness is widespread in the cluster core 
and much of it does not correlate with radio emission or jet axis; hence it
could arise from bubbles of thermal plasma ejected from the AGN.

Initially, the radio jets will propagate within the region of the bubble 
(the wind zone). However, since the leading edges of the jets can advance 
much faster than the bubble's expansion, the resulting hotspots would soon 
emerge out of the bubble.  The relativistic plasma back-flowing from them 
would now begin to ``impinge" on the bubble's surface from outside. Although 
a majority of radio lobes seem to be roughly pressure matched with the ambient 
medium into which they are expanding (e.g., Hardcastle \& Worrall 2000; Hardcastle et al.\ 
2002; Croston et al.\ 2004a), over-pressured lobes also exist.  For 
instance, such a situation has been found even for highly expanded radio lobes 
associated with giant radio galaxies (Kronberg et al.\ 2001). In the framework 
of our model, it is such radio sources that are more conducive to the formation 
of superdisks. Although the lobes can be significantly overpressured, the bubble 
is typically only marginally so, and  hence the lobes can be expected to prevent 
the expansion of the wind bubble in their direction (i.e., along the radio axis).
Taking into account the backflow within the lobes one can expect even 
compression of the wind bubble along the radio axis. Such a sustained squeezing 
from two opposite sides could transform the wind bubble into a fat disk. This 
would sometimes appear as a ``chimney'' when projected on the sky.  This 
process of sideways compression or constriction by the lobe pair would deflect 
and accelerate the outflowing AGN wind roughly normal to the jet direction, 
through the putative fat disk. Note that in this picture, the boundaries of
the radio gaps are identified as  {\it lobe-wind interfaces}.

Unfortunately, radio maps of good enough quality to convincingly reveal 
a superdisk (SD) morphology are presently available only for relatively
nearby RGs ($z < 0.6$), and for them there is a high probability of substantial 
amounts of circum-galactic hot gas being present that could confine the bubble
and even the radio lobes. 
Obtaining good evidence for SDs from radio maps of galaxies at 
the higher redshifts for which external hot gas is  less likely to be important is quite difficult.  Such radio maps require a combination
of high angular resolution, excellent sensitivity and high dynamic range, and this is  particularly hard to achieve because of the $(1+z)^{-4}$ dimming of surface brightness. In addition, these observations would preferably be made at lower 
frequencies where morphologies of the lobes are best discerned. 
Nonetheless, a small number of somewhat higher redshift candidate SDs can be located in the literature, e.g.:  0838$+$133 ($z = 0.684$; Sambruna et al.\ 2004); 3C 265 ($z = 0.811$; Bondi et al.\ 2004);
1040$+$123 ($z=1.029$; Sambruna et al.\ 2004).
 
Additional evidence for SDs in high-$z$ RGs comes, at present, from optical 
detections of the strong Ly$\alpha$ emission associated with the RGs.  This 
often very extended Ly$\alpha$ emission traces any dusty screen of gas because 
resonant scattering greatly enhances the absorption of these photons by dust 
(e.g., GKW00).
All the radio-loud objects with two-dimensional Ly$\alpha$ images showing 
substantial asymmetries of which we are aware are, in order of increasing 
distance: 3C 294 ($z = 1.82$; McCarthy et al.\ 1990); 3C 326.1 ($z = 1.82$; 
McCarthy et al.\ 1987); 1138$-$262 ($z=2.16$; Pentericci et al.\ 1997); 0200$+$015 ($z=2.23$; Jarvis et al. 2003)
0852$+$124; ($z=2.47$; Gopal-Krishna et al.\ 1995); B3 J2330$+$3927 ($z=3.09$; 
P{\'e}rez-Torres et al.\ 2006); MRC 0316$-$257 ($z=3.13$; 
Venemans et al.\ 2005); 6C 1232$+$39 ($z=3.22$; Eales et al.\ 1993); 1243$+$036 
($z=3.57$; van Oijk et al.\ 1996); 4C 41.17 ($z=3.80$; Chambers, Miley \& van 
Breugel 1990); and 8C 1435$+$635 ($z=4.26$; Spinrad, Dey \& Graham 1995).
For six of the seven RGs from this list for which the orientation of the radio 
lobes can be determined, the Ly$\alpha$ asymmetry is in the sense that stronger 
emission is seen from the side of the approaching lobe.  
Thus, the most likely explanation is that the Ly$\alpha$ photons coming from the
farther side of the active nucleus are being obscured by a large gas disk around the parent galaxy (GKW00).

\section{Jet--Bubble Interactions}

\subsection{Wind Blown Bubbles}
Levine \& Gnedin (2005) have recently modelled expansion of hot thermal wind 
outflowing from an AGN. In this scenario they combine cosmological simulations 
with an AGN luminosity function and place the AGN at local IGM density 
peaks.  They find that large portions of the IGM can be filled with such 
outflows unless kinetic luminosities are $< 10\%$ of the bolometric 
luminosities. This remarkable conclusion is similar to that reached by 
Furlanetto \& Loeb (2001) for similar spherical outflows and by Gopal-Krishna 
\& Wiita (2001) and Kronberg et al.\ (2001) for outflows in the form 
of relativistic jets (cf.\ Barai \& Wiita 2007).  Therefore, the impact of 
AGNs on the formation of structure in the universe may be quite significant,
by triggering/accelerating global star formation in a multiphase intergalactic
medium (e.g., De Young 1989; Rees 1989; Choskhi 1997; Gopal-Krishna \& Wiita 
2001; Gopal-Krishna, Wiita \& Osterman 2003; Silk 2005; Barai \& Wiita 2006, 
2007) or inhibiting it in the case of a single-phase IGM (e.g., Rawlings \& 
Jarvis 2004; Scannapieco, Silk \& Bouwens 2005).  The role of large radio lobes 
in spreading magnetic fields (Kronberg et al.\ 2001; Gopal-Krishna \& Wiita 
2001; Gopal-Krishna et al.\ 2003; Barai et al.\ 2004) and in distributing metals (Gopal-Krishna 
\& Wiita 2003; Gopal-Krishna, Wiita \& Barai 2004) through the IGM may also 
be very significant.

The Levine \& Gnedin (2005, hereafter LG) model builds upon that of Scannapieco 
\& Oh (2004, hereafter SO) and we shall employ results from each of these 
papers in the following. Assume that the AGN injects thermal wind with kinetic
energy 
$E_w = L_w \tau_{\rm act}$, where
$L_w (\equiv \epsilon_w L_{bol})$ is the kinetic luminosity, 
$\tau_{\rm act}$ is the time the AGN is active in this
mode and $L_{\rm bol}$, the bolometric luminosity, is a measure of the accretion power.  Then,
if $\delta_m$ is the mean relative overdensity in the environment in which the AGN
is situated (assumed to be constant over the relevant volume) then the radius of the
(spherical) bubble inflated, for a time $t > \tau_{\rm act}$ will be (LG)
\begin{equation}
R_S = 1.7~{\rm Mpc} ~E_{w,60}^{1/5} (1+\delta_m)^{-1/5} (1+z)^{-3/5} t_{9}^{2/5},
\end{equation}
where $E_{w,60} = E_w/10^{60}{\rm erg}$, $z$ is the redshift and $t_{9}$ is the time since
the active phase began, in units of Gyr.
The bubble should grow according to Eq.\ (1) until it reaches pressure equilibrium with the
ambient environment, at which point the size of the bubble would be (LG)
\begin{equation}
R_P = 3.24 \times 10^{-5} {\rm Mpc} \Bigl( \frac{3E_{w,60}}{4\pi P} \Bigr)^{1/3},
\end{equation}
where the external thermal pressure due to the IGM, $P = (1+\delta_m) n_b k_B T$; here its mean baryon density, $n_b =  
3.3\times 10^{-7}(1+z)^3~{\rm cm}^{-3}$, and the typical IGM temperature, which 
is taken to be constant with epoch, is $T \simeq 1.5 \times 10^4$K (LG).
We note that it might be more appropriate to allow for the average
IGM temperature to increase as
$z$ decreases (Cen \& Ostriker 1999), but including this effect does not make
significant changes in our argument, whereas considering expansion into a much 
hotter and significantly denser ICM would do so, in that the radii out to 
which the bubbles would expand supersonically would be much smaller (e.g., 
Sternberg et al.\ 2007).

At times after the wind energy has been completely injected, the velocity of 
the bubble's expansion is given by (SO)
\begin{equation}
v_S = 1470 ~{\rm km ~s}^{-1} E_{w,60}^{1/2} R_{S,{\rm Mpc}}^{-3/2} (1+\delta_m)^{-1/2} (1+z)^{-3/2},
\end{equation}
where $R_{S,{\rm Mpc}}$ is $R_S$ in units of Mpc.  This result assumes that 
$\tau_{\rm act}$ is short compared the cooling time for the hot gas in the 
bubble so that radiative losses can be ignored; in addition, $PdV$ work and the 
work done against the galaxy's gravity can also be ignored (SO, LG).  These are 
all plausible approximations  (e.g., Furlanetto \& Loeb 2001). The final 
approximation involved in these formulae is that the inertia of the displaced 
galactic ISM is less than that of the IGM, which will be reasonable
once $R_S$ exceeds about 10 kpc. 

We find that the corresponding equation for earlier times, $t < \tau_{\rm act}$, is  
\begin{equation}
R_S = 1.17  (\epsilon_w L_{bol})^{1/5} \rho_{\rm IGM}^{-1/5} t^{3/5}, 
\end{equation}
where $L_w = \epsilon_w L_{\rm bol}$, with $L_{\rm bol}$ the bolometric luminosity, and
$\rho_{\rm IGM}(z) =  (1+z)^3 (1+\delta_m) \rho_{\rm IGM}(0)$.  In Eq.\ (4) and in all other expressions where the unit of the result is not specified,
we are employing cgs units.  At these earlier times we obtain
\begin{displaymath}
v_S = 3060 ~{\rm km ~s}^{-1} \Bigl( \frac{\epsilon_w L_{\rm bol}}{3.17 \times 10^{44} {\rm erg ~s}^{-1}} \Bigr)^{1/3} R_{S,{\rm Mpc}}^{-2/3} 
\end{displaymath}
\begin{equation}
~~~~~~~~~~~~(1+\delta_m)^{-1/3}  (1+z)^{-1}.
\end{equation}
Note that one may reasonably expect that $L_w \le L_B \sim 0.1 L_{\rm bol}$, 
with $L_B$ the B-band luminosity (LG; cf.\ Rawlings 2003). Therefore, values of
$\epsilon_w$ between 0.01 and 0.1 are reasonable, with SO arguing for 
$\epsilon_w \approx 0.05$, and most of our computations assume such a value.
However, other authors have argued that the power in the mechanical outflow can 
substantially exceed $L_{\rm bol}$
(Churazov et al.\ 2002, 2005; Peterson \& Fabian 2006); this situation may be 
generally the case for radiatively inefficient 
accretion onto black holes (e.g., Hopkins, Narayan \& Hernquist 2006).

The outflowing wind carries enough momentum to carve out a low density cavity or bubble, but it is also depositing some mass into that bubble.
To provide an estimate of the density of this bubble through which the jets later will be propagating in our scenario, 
we must introduce another parameter, which can be taken to be the wind speed, $v_w$.  Then we have an expression for the rate at which mass is ejected in the non-relativistic wind
$\dot{M}_w$
\begin{equation}
\dot{M}_w = 2L_w  v_w^{-2} = 6.3 \times 10^{27}{\rm ~g~s}^{-1} v_{w,4}^{-2} E_{w,60} \tau_8^{-1},
\end{equation}
with $v_{w,4} = v_w/(10^4 {\rm km ~s}^{-1})$ and $\tau_8 = \tau_{\rm act}/10^8 {\rm yr}$.
A plausible, but by no means rigid, constraint on the rate at which mass outflows in this wind is that it should be less than the mass accretion rate
needed to power the AGN.  Using the typical parametrization for an standard (radiatively efficient)  accretion disk, where $L_{\rm bol} = \epsilon_{\rm acc}    \dot{M}_{\rm acc} c^2$, where $\epsilon_{\rm acc} \sim 0.1$, we find that if we adopt
the above constraint
\begin{equation}
v_w > (2 \epsilon_{k} \epsilon_{\rm acc})^{1/2} c = 0.045 c~ \epsilon_{k,-2}^{1/2} \epsilon_{{\rm acc},-1}^{1/2}.
\end{equation}
Thus the minimum wind speed for a radiatively efficient accretion flow is, $v_{w,\rm min} \simeq 10^4 {\rm km ~s}^{-1}$, where we have evaluated Eq.\ (7) taking $\epsilon_{k} = 0.01$ and
$\epsilon_{\rm acc} = 0.057$, the latter being the minimum value for thin disk accretion onto a non-rotating black hole (e.g. Shakura \& Sunyaev 1973).  It is, however, important to note that substantially lower wind speeds are allowed if the accretion rate is sufficiently low so that a radiatively inefficient accretion flow (e.g., Narayan \& Yi 1994) is established.  It is under these circumstances that
$L_w$ can exceed $L_{\rm bol}$.

Since the total mass ejected in the wind up to time $t < \tau_{\rm act} $ is  $\dot{M}_w t$, the
mean density in a wind-filled bubble is
\begin{displaymath}
\rho_b(t) = 3\dot{M}_w t/4 \pi R_S^3(t)  
\end{displaymath}
\begin{equation}
= 0.298 ~(L_{\rm bol} \epsilon_w)^{2/5} t^{-4/5} \rho_{\rm IGM}^{3/5}(0) (1+z)^{9/5} (1+\delta_m)^{3/5} v_w^{-2},
\end{equation}
where we have used Eq.\ (4).

However, once $t > \tau_{\rm act} $, $R_S$ is better approximated by Eq.\ (1) and no more mass is injected, giving $M_w =  \dot{M}_w \tau_{\rm act} = 2E_w / v_w^2$ and thus
\begin{displaymath}
\rho_b(t) = 3 M_w /4 \pi R_S^3(t) 
\end{displaymath}
\begin{equation}
~~~= 0.298 E_w^{2/5} t^{-6/5} \rho_{\rm IGM}^{3/5}(0) (1+z)^{9/5} (1+\delta_m)^{3/5} v_w^{-2}.
\end{equation}
So the bubble density declines faster after a time $\tau_{\rm act}$ 
($\propto t^{-6/5}$) than it did
before that time ($\propto t^{-4/5}$) even though the radius expands more slowly for $t > \tau_{\rm act}$.  This gas density is usually extremely low, but the temperature of this gas is typically hot enough to produce
some X-ray emission (e.g., Krause 2005).

\subsection{Jets Launched into the Bubble}
We now assume that at time $t_j$, measured from the onset of the wind
 outflow, a pair of jets
of equal kinetic power 
is launched; the total
power in the jets is $L_j$.  We will further assume that $L_j$ remains constant 
throughout the duration of the jets' ejection,  $\tau_j$.
Because the jets are both focussed and ploughing through a low density medium in the wind-swept bubble, they will propagate outward substantially faster than would a quasi-spherical
wind of similar power.  Two-dimensional hydrodynamic simulations of a particular case of a pair of very powerful jets propagating through a 
supersonic bubble produced by a supernova driven galactic wind have recently been performed by Krause (2005).
These computations were designed to explain the clumpy HI absorption seen in many high-$z$ RGs in terms of instabilities produced in the wind shell upon penetration by the jet. They clearly show that a powerful jet launched a long time ($t_j =$ 80 Myr in this case) after the wind initiation will, as we assume, produce a narrow cocoon within the bubble. Then, after passing though the bubble wall, the jet will continue to propagate more slowly in a denser ambient medium, but will still produce extended overpressured lobes (Krause 2005).   These simulations neglected magnetic
field in the bubble, whereas we expect the bubble to become magnetized,
even if it was not originally so,
due to some mixing with the synchrotron plasma of the radio lobes blown by 
the jets during their passage through the bubble. Any magnetization should
enhance the bubble's stability against surface distortions (e.g., Kaiser et al.\ 2005; Lyutikov 2006).

Following the basic arguments of Kaiser \& Alexander (1997) and calibrating the 
models through catalogs of powerful radio galaxies (Blundell, Rawlings \& 
Willott 1999; Barai \& Wiita 2006, 2007) we have a lower bound on the distance 
out to which a jet of average collimation will have propagated within the
bubble, at time $t > t_j$,
\begin{equation}
R_j(t) \geq 1.57  (t-t_j)^{3/5} L_j^{1/5} \rho_b^{-1/5}(t_j),
\end{equation}

as long as we ignore the spatial gradient in density within the bubble, which 
is an approximation adequate for our purposes. The above expression could be 
an equality if $\rho_b$ did not continue to decline after $t_j$; however, if 
the velocity at which the jet moves is sufficiently high compared to the 
rate of expansion of the bubble after the launch of the jet (as usually will be 
the case in the situations of interest to us), then this continuing temporal 
decline of $\rho_b$ has only a small effect, which we shall ignore.  Hence the 
numerical results are computed assuming Eq.\ (10) is an equality.  Only if 
$t_j \ll \tau_{\rm act}$ and $\dot{M_w}$ is large do we have such high values 
of $\rho_b(t_j)$ that the jets will have inordinate difficulty in catching up 
to the wind under the approximation we employ; under these circumstances the 
continued decline in $\rho_b$ with time really should be considered.
In the following exploratory calculation we also ignore the slowing down the 
jet would suffer while penetrating the reverse shock and the denser shell of 
material swept up by the expanding bubble. In addition, we ignore the modest 
acceleration produced in the boundary  of the bubble through the extra pressure
injected by the jet. We note that this slowing of the jet's relative speed with 
respect to that of the wind arising from these effects is roughly offset by the 
relative acceleration produced by the continuing temporal decline of the bubble 
density.

\begin{figure}
\hskip -0.8cm
\psfig{file=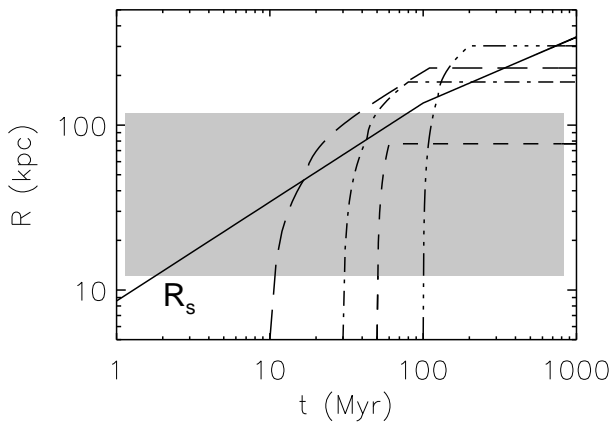}
\hskip 0.0cm
\caption{Radii of a wind-blown bubble, $R_S$, and of four jets launched within that bubble.  The solid curve shows $R_S$ for the following parameters: $L_w = 5\times 10^{42} {\rm erg~s}^{-1}$, $\tau_{\rm act} = 100~{\rm Myr}$, $z=1.0$, $\delta_m = 5.0$, $v_w = 3000~{\rm km~s}^{-1}$.
The other curves are propagation distances for the jets, all of which have $L_j = 5\times 10^{42}{\rm erg~s}^{-1}$.  Jet parameters are: $t_j = 10~{\rm Myr}$ and  $\tau_j = 100~{\rm Myr}$ (long-dashed); $t_j = 30~{\rm Myr}$ and  $\tau_j = 50~{\rm Myr}$ (dot--dashed);
$t_j = 50~{\rm Myr}$ and  $\tau_j = 10~{\rm Myr}$ (short-dashed); $t_j = 100 ~{\rm Myr}$ and  $\tau_j = 100~{\rm Myr}$ (triple-dot--dashed).  The grey band shows the plausible values of $R_{eq}$, with lobe magnetic fields of $3.0 \times 10^{-5}$ gauss and $1.0 \times 10^{-6}$ gauss corresponding to its lower and upper boundaries, respectively.}
\end{figure}

In Fig.\ 1 we display the expansion of both the wind-blown bubble and the jets
launched later for four representative combinations of parameters.  The decline 
of the bubble expansion rate is seen at 100 Myr. In one illustrated case the jet
never catches up to the wind's boundary since this jet is launched 50 Myr after 
the wind and is, moreover, energised for only 10 Myr.  In the other three cases, the jets do catch up to the bubble boundary, despite having a shorter lifetime 
(in one case) and lower total energies (in two cases).  Once the jets pass 
through the bubble and enter the IGM their velocities of advance are seen to 
change.  With our assumptions that the bubble continues to inflate even after 
the AGN wind is shut off, while the jet-fed cocoons stop growing right after 
the jet is switched off, we see that in all three of these displayed cases the
bubble boundary could eventually once again overtake the jet; however, thanks to the confinement exerted by the cocoon on the bubble in the interim, the times 
when this occurs that can be read from Fig.\ 1 are underestimates.

We can now define the catch-up time, $t_c$, and catch-up radius, $R_c$, through 
the relation $R_j(t_c) = R_S(t_c) \equiv R_c$.  The formula in Eq.\ (10) is 
valid for $t_j < t < t_c$, but at later times the jets are ploughing through 
the constant density IGM, and we have for $t_c < t < t_j+\tau_j$
\begin{equation}
R_j(t) = R_c + 1.57 (t-t_c)^{3/5} L_j^{1/5} \rho_{\rm IGM}^{-1/5}(z).
\end{equation}

\begin{figure*}
\psfig{file=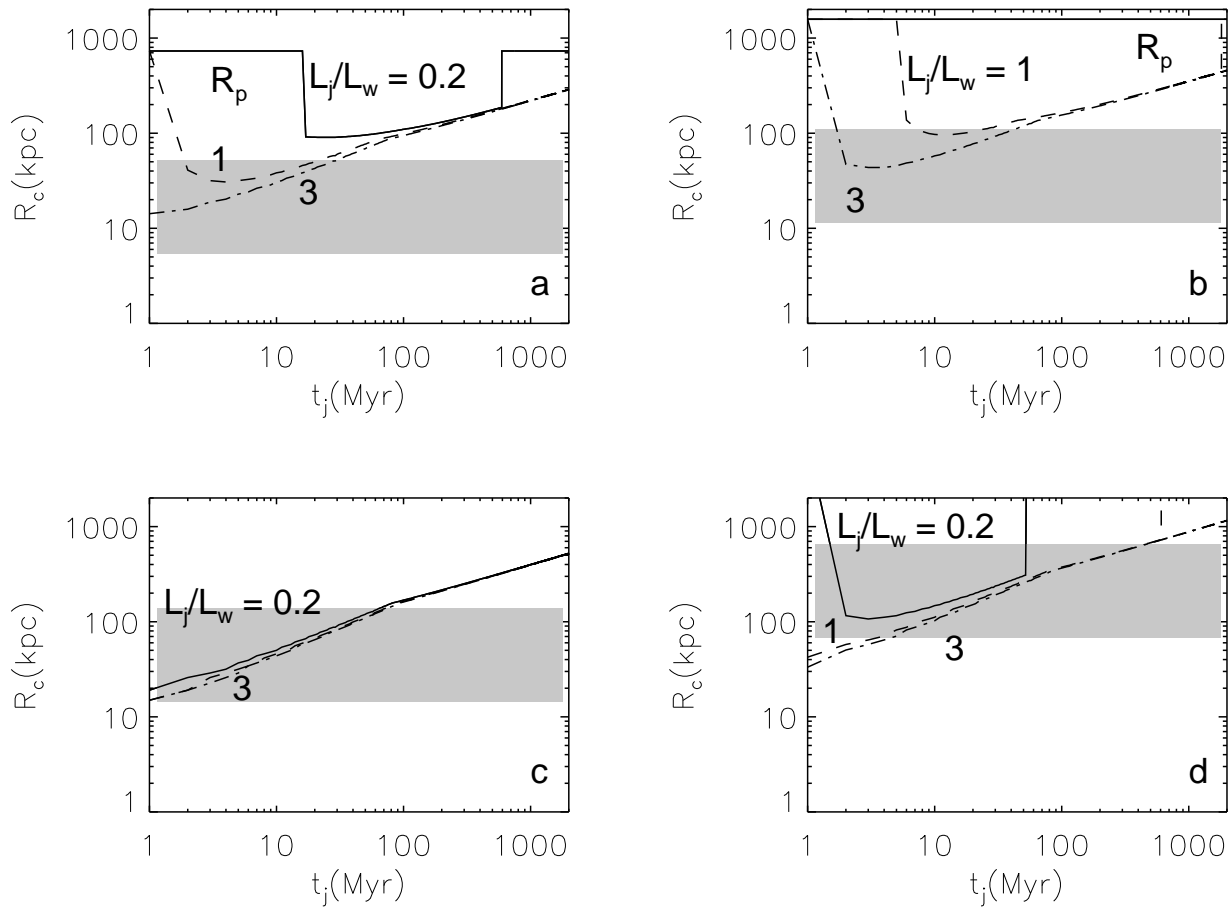} 
\caption {Catch-up radii, $R_c$, as functions of $t_j$, shown for three different ratios of the jet to wind luminosities: 0.2 (solid), 1.0 (dashed), 3.0 (dot-dashed).  All panels share the following parameters, $z = 1$ and $\tau_{\rm act} = 1.0 \times 10^8 {\rm yr}$, and a)--c) also share the following parameters, $\delta_m =5$, $\epsilon_w = 0.05$, and $\tau_j = 1.0 \times 10^8 {\rm yr}$.  The other parameters are: a) $L_w = 5.0 \times 10^{41} {\rm erg s}^{-1}$, $v_w = 1.0 \times 10^3 {\rm km~s}^{-1}$; b) $L_w = 5.0 \times 10^{42} {\rm erg~s}^{-1}$, $v_w = 1.0 \times 10^3 {\rm km~s}^{-1}$; c) $L_w = 1.0 \times 10^{43} {\rm erg~s}^{-1}$, $v_w = 1.0 \times 10^4 {\rm km~s}^{-1}$; d) $L_w = 1.0 \times 10^{45} {\rm erg~s}^{-1}$, $v_w = 1.0 \times 10^4  {\rm km ~s}^{-1}$, and now $\delta_m =10$, $\epsilon_w = 0.10$ and
$\tau_j = 3.0 \times 10^7 {\rm yr}$.  As in Fig.\ 1, the shaded areas delineate values of $R_{eq}$ for lobe magnetic fields between 30 $\mu$G (lower boundary) and 1 $\mu$G (upper boundary); $R_P$ is shown
in a) and b).}
\end{figure*}

In Fig.\ 2 values for $R_c$ are plotted against $t_j$, the 
time interval between the wind's onset and the jets' launching, for substantial 
ranges of the most important input parameters. Each of the panels shows the 
results for two or three different jet powers. The values of $R_P$ (imposed by 
the IGM; Eq.\ 2) are shown in Figs.\ 2a and 2b for $E_w = 1.58 \times 10^{57}$ 
erg and $E_w = 1.58 \times 10^{58}$ erg, respectively, while the corresponding 
values of $R_P$ for $E_w = 3.16 \times 10^{58}$ erg (Fig.\ 2c) and 
$E_w = 3.16 \times 10^{60}$ erg (Fig.\ 2d) exceed 2 Mpc and are not shown. 
The convention used in Fig.\ 2 is that if the jet never catches up to the 
bubble's boundary then $R_c$ is plotted as  $R_p$.  

Under most circumstances we see that $R_c$ rises as $t_j$ increases, as 
expected, since a jet should take longer to catch up to a bubble which has had 
a greater head-start. In all cases, more powerful jets obviously produce lower 
values of $R_c$ for fixed values of $t_j$.  The less powerful jets (e.g., 
$L_j/L_w = 0.2$ in Figs.\ 2a, 2b and 2d; $L_j/L_w = 1.0$ in Figs.\ 2b and 2d), 
and even very powerful, but short-lived, jets ($L_j/L_w = 3.0$ in Fig.\ 2d) may 
never catch up to the wind boundary if the jet's launch is sufficiently delayed.  
The assumption that $\rho_{b}$ is fixed at the value it has attained at $t_j$ 
leads to the peculiar results plotted at low values of $t_j$ in Figs.\ 2a, 2b, 
and 2d for some of the adopted values of jet power. Because these nominal 
densities can be quite high for small values of $t_j$ (see Eq.\ 8) the jets 
appear to never catch up to the wind boundary; since, in actuality the bubble 
density would continue to decline, unless the jets were very short-lived they 
would indeed eventually catch up to the wind's boundary.

We shall now briefly examine some broad implications of the above analytical 
scheme, taking a fiducial values of $E_w \sim 10^{59}$ erg for the kinetic 
energy of the wind outflow and the same value for the kinetic energies of
the relativistic jets. From Eq.\ (1) and Fig.\ 2, it is seen that if the 
bubble's expansion is allowed to continue much after cessation of the wind 
outflow ($\tau_{\rm act} \approx 10^8$ yr), then in the course of $\sim 10^9$  
years it would have grown to a quasi-spherical volume of size around 1 Mpc if 
it  encounters an ambient IGM pressure  $< 10^{-16}$ dyne cm$^{-2}$; this will 
be the case provided the AGN is located the field, rather than inside a dense 
cluster environment, and is not at too high a redshift so that the combination
$\delta_m (1+z)^3 < 150$.

\begin{figure}
\caption{An illustration of three stages in the production of 
superdisk through the initial inflation of a wind-blown bubble to the
recompression of the bubble by the radio lobes after the jets have penetrated 
through it. [Submited as a .jpg file.]}
\end{figure}

However, if a relativistic jet is subsequently launched, then once it crosses 
the bubble's boundary, the much higher pressure of its radio lobe would begin 
to act as a wall, blocking, and even reversing the bubble's expansion along 
the jet's direction.  The cartoon in Fig.\ 3 shows this as the last phase of 
this dual mode AGN energy output.  

The pressure in the radio lobe is indeed much higher than the IGM pressure, 
since it is given by $P_{\rm lobe} \simeq 8 \times 10^{-14} B^2_{-6}$ dyne 
cm$^{-2}$, assuming the minimum energy condition which signifies an energy 
equipartition between magnetic and relativistic particles; here $B_{-6}$ is 
the lobe field in microgauss and values of $B_{-6}$ between 1 and 30 are 
typically estimated for FR II RGs.  Inserting this value of $P_{\rm lobe}$ 
into Eq.\ (2) yields an equilibrium radius of the bubble, $R_{eq} \simeq 
0.22~{\rm Mpc}~ E^{1/3}_{w,59} B^{-2/3}_{-6}$, which is in the range from 
a few to hundreds of kiloparsecs.  These values are also displayed in 
Figs.\ 1 and 2 as grey bands corresponding to the range of values of $B$ 
mentioned above. 

Note that the higher pressure in the radio lobes may even reverse the expansion 
of the bubble boundary in the areas of contact with the lobes if $R_c >
R_{eq}$. Thus, the actual width of the gap between the radio lobes should be 
between $2R_{eq}$ and $2R_c$, assuming that the jets live long enough and are 
launched soon enough to overtake the bubble. The consistency of these estimates 
with the observed widths of the radio gaps in SDs argues for the efficacy 
of the wind's confinement by the lobes, as envisioned in our model.  We can 
estimate the time scale over which this recompression along the jet axis from 
$R_c$ to $R_{eq}$ could occur as 
\begin{displaymath}
\Delta t \simeq 2.5 \times 10^7 {\rm yr} \Bigl( \frac{R_c}{100 {\rm kpc}} \Bigr)^{1/2}
 \Bigl( \frac{R_c-R_{eq}}{100 {\rm kpc}} \Bigr)^{1/2}
\end{displaymath}
\begin{equation}
~~~~~~~~~~~~~~~~~ \times (1+z)^{3/2}(1+\delta_m)^{1/2}B^{-1}_{-6}.
\end{equation}
As long as $\tau_j > t_c + \Delta t$ this recompression is expected to be 
important and the half-width of the superdisk is likely to be close to $R_{eq}$. 
 It should be noted that the compression would probably be enhanced by a dynamic
pressure arising from the back-flow in the radio lobes, which has been found to 
be present in simulations of radio jets (e.g., Hooda \& Wiita 1998). 
Nonetheless, $R_c$ provides an upper limit to this half-thickness under most 
circumstances where both winds and jets are important carriers of AGN output. 


If the value of $t_j$ is so small and the jet luminosity is so large that 
$R_c < R_{eq}$, then the pressure inside the bubble may still be greater than 
that inside the lobes for some values of $t > t_c$.  In this case the bubble 
may continue to expand in all directions until $R_S \simeq R_{eq}$, though a 
somewhat smaller maximum value along the jet directions is expected because 
of the additional ram pressure exerted by the backflow in the lobe.  So again, 
the width of any observed superdisk should be between  $2R_{eq}$ and $2R_c$.


Beyond this stage, the subsequent expansion of the bubble's plasma would be 
confined to a plane roughly orthogonal to the jets' direction, giving rise to 
a pancake-like region filled with the hot outflowing wind material. 
Of course, as mentioned above, in the extreme cases where the bubble's
confinement by the overtaking lobes begins to take place after a very long 
time ($t_c \sim 10^9$ yr), which is conceivable for giant radio galaxies, the 
radio gap could approach megaparsec dimensions, as seen for the radio galaxy 
0114$-$476 (Saripalli et al. 2002).  

\section{Discussion and Conclusions}

If AGN activity usually involves both quasi-spherical wind outflows and the 
subsequent launching of collimated jets the latter will frequently overtake 
the former and can produce  large gaps between the radio lobe pair, particularly in 
the higher redshift RGs which are yet to acquire a substantial, high-pressure 
circumgalactic medium.  Here we have employed a highly simplified model to 
estimate some of the key properties of these wind/jet interactions.

It is important to  recall that we have made several assumptions and 
approximations, most of which were discussed in Sect.\ 3:
(i) spherical symmetry of the wind; (ii) neglect of the mass of the host 
galaxy's ISM which is mostly swept into a shell; (iii) neglect of the reverse 
shock behind the wind; (iv) neglect of work done against the galactic 
gravitational potential and of $PdV$ work done by the expanding bubble;
(v) neglect of the likely correlated growth in overdensity and temperature in 
the filamentary IGM; (vi) neglect of the decline with time of 
$\rho_{b}$ for $t_j < t < t_c$; (vii) neglect of the additional pressure inside 
the lobes for $t_j < t  < t_c$; (viii) ignoring the possibility that for
$t_c < t < \tau_{\rm act}$ the continuing outward wind flow could destabilize 
the jet (Norman, Burns \& Sulkanen 1988); (ix) assuming that the bubble 
continues to grow after $\tau_{\rm act}$, while cutting the growth of the jets 
immediately after $t_j + \tau_j$.  Removing approximations (iii) and (vii) 
would tend to increase $R_c$ and $t_c$; however, eliminating approximations 
(iv), (v) and (vi) would decrease those quantities, and hence typically narrow 
the observed superdisks.  Approximation (ii) reduces $R_S(t)$ but it also 
reduces $R_j(t)$ and so its effect on $R_c$ is unclear.  
The early destabilization of a not very supersonic jet by a very supersonic 
wind (removing approximation viii) would not change $R_c$, but would reduce
the chance of bubble compression.  Allowing for the relaxation of assumption 
(ix) can lower $R_c$ and has the opposite effect on continuing compression. 
Clearly,  a more appropriate way to treat this problem would be through detailed 
hydrodynamical simulations which would allow the relaxation of all of the above 
assumptions, but given the countervailing tendencies noted above, we feel 
confident that our basic results are  quite plausible, although no numbers 
we provide can be considered precise. We have already emphasized that the  
IGM conditions we consider are appropriate for powerful (FR II) RGs forming 
during the peak of the radio luminosity function at $1 < z < 3$ (e.g., Grimes, 
Rawlings \& Willott 2004). In GKW00 we provided indirect evidence for 
superdisks in Ly$\alpha$ galaxies in and above this redshift range; however, 
our  explicitly mapped examples of RGs with superdisks must be drawn  mostly from 
closer objects for which radio maps with adequate resolution and dynamic
range  are available, and for which the actual conditions are often
more similar to those of the ICM at lower redshifts. Under those circumstances 
the typical sizes for wind inflated bubbles would be compressed to a few kpc. 

We now summarize some salient features of our model.

(1) Superdisks (at least the wider ones) mark {\it giant}  planar conduits 
along which the AGN's non-relativistic wind is able to escape preferentially, 
i.e., in directions roughly perpendicular to the radio axis along which 
synchrotron plasma is transported. 
We suggest that the surface of dynamical interaction between the wind bubble 
and the radio lobe pair can often give rise to well-defined, planar boundaries 
between the two media, which we identify as {\it lobe-wind interfaces} and 
which  may be manifested in the RGs  as central emission gaps or 
superdisks. (In most RGs, however, the viewing angle would work against the
sharp edges, i.e., a superdisk morphology, being observed). The metal enriched 
gas swept out of the host galaxy is thus transported to great distances, not 
only by being dragged along by the jets, but in the perpendicular direction 
as well (through the superdisk). Thus, the AGN activity would tend to 
isotropize the metal enrichment process, as is indeed found from measurements 
(de Grandi et al.\ 2004).

(2)  This scenario provokes us to revisit the question raised three decades 
ago by Jenkins \& Scheuer (1976):  ``what docks the tails of radio source 
components?" They concluded that the cause is other than synchrotron losses. 
Later, an explanation was proposed in terms of blocking of the radio lobe 
plasma by the ISM of the host galaxy (e.g., Leahy \& Williams 1984). 
In our picture, on the other hand, the sharp and straight edges of the 
strip-like central gaps in the radio bridge signify {\it active surfaces} 
where the thermal wind outflowing from the AGN is actually countered and 
redirected by the lobe overpressure assisted by the dynamic pressure of the 
back-flowing relativistic plasma inside the radio lobes. Eventually, when the
pressure of this back-flow has declined sufficiently, it could even mix with 
and be dragged along the thermal wind escaping in the perpendicular direction.

(3)  If indeed the ``wind" outflow precedes the radio jet ejection, the
pancake or superdisk resulting from the lobe-wind interaction can become 
``frozen" in space quite early in the active phase of the galaxy. Now, if the 
galaxy has a large enough component of motion normal to the superdisk  (say, 
500 km s$^{-1}$ or more), it may even move out of the latter into the lobe 
during its active lifetime.  Two RGs exemplifying such a situation are 
3C~16 and 3C~19, where the host galaxy is seen {\it outside} the radio gap
(Harvanek \& Hardcastle 1998; Leahy \& Perley 1991; Gilbert et al.\ 2004). 
While such a morphology is expected only in extreme cases, it is quite 
conceivable in our picture but hard to understand within the usual 
interpretation of the radio gaps invoking a buoyancy led outward squeezing 
of the radio lobes by the denser ISM of the parent galaxy.
 
(4) Notwithstanding the sharp, quasi-planar  boundaries of the radio gap, the 
present model  does allow for some fainter radio emission seen within the 
gap (being remnant of the early phase when the radio jets were still boring 
their way through the  bubble). Examples of this can be found in the radio 
sources J1137$+$613 (Lara et al.\ 2001) and J1628$+$3932 (de Breuck et al.\ 
2004).  Otherpossible manifestations of this situation are 3C~63 (Harvanek \& 
Hardcastle 1998),  3C~136.1 (Leahy \& Williams 
1984), and 3C~300 (Leahy \& Williams 1984; Hardcastle et al.\ 1997).
However, such remnant emission usually will be difficult to see without high 
dynamic range observations and will not last very long compared to the total 
lifetime of the radio source.  This is because this inner radio emitting plasma 
will be cut off from a continued supply of the backflow from the lobes and will 
be mixed up and dispersed with the the thermal outflow through the chimney 
(i.e., superdisk).

Finally, although the scenario sketched here has been quantified in a 
highly simplified analytical form, the possibility of the sharp-edged radio 
gaps being  wind-lobe interfaces, i.e., ``active" surfaces of dynamical 
interaction between the thermal and nonthermal outflows, may have other 
interesting observational and theoretical consequences. Hence, this general 
picture needs to be explored further.  Full hydrodynamic simulations of
this situation are worth pursuing. On the observational side, detailed radio 
imaging of high$-z$ radio galaxies would provide useful input and constraints 
on the basic model presented here.

\section*{Acknowledgments}

We thank the referee, Martin Hardcastle, for criticisms which significantly
focussed the arguments of this paper.
This research has made use of the NASA/IPAC Extragalactic Database (NED)
which is operated by the Jet Propulsion Laboratory, under contract with
the National Aeronautics and Space Administration. P.J.W.\ is grateful for 
continuing hospitality at the Princeton University Department of Astrophysical
Sciences and acknowledges support from a sub-contract to GSU from National 
Science Foundation grant AST-0507529 to the University of Washington.


\begin{thebibliography}{99}
\bibitem{}Alexander P., 1987, MNRAS, 225, 77
\bibitem{}Barai P., Wiita P.\ J., 2006, MNRAS, 372, 381
\bibitem{}Barai P., Wiita P.\ J., 2007, ApJ, 658, 217
\bibitem{}Barai P., Gopal-Krishna, Osterman M.\ A., Wiita P.\ J., 2004, BASI, 32, 385
\bibitem{}Binette L, Wilman R.\ J., Villar-Martin M., Fosbury R.\ 
A.\ E., Jarvis M.\ J., R{\"o}ttering, H.\ J.\ A., 2006, A\&A, 459, 31
\bibitem{}Binney J. 2004, in Reiprich T., Kempner J., Soker N., eds, The Riddle of Cooling Flows in Galaxies and Clusters of Galaxies, published electronically at www.astro.virginia.edu/coolflow/proc.php
\bibitem{}Black A.\ R.\ S., Baum S.\ A., Leahy J.\ P., Perley R.\ A., Riley J.\ M., Scheuer P.\ A.\ G., 1992, MRNAS, 256, 186
\bibitem{}Blandford R.\ D., Begelman M.\ C., 1999, MNRAS, 303, L1
\bibitem{}Blandford R.\ D., Znajek R.\ L., 1977, MNRAS, 179, 433
\bibitem{}Blundell K.\ M., Rawlings S., Willott C.\ J., 1999, AJ, 117, 677
\bibitem{}Bogers W.\ J., Hes R., Barthel P.\ D., Zensus J.\ A., 1994, A\&AS,
105, 91
\bibitem{}Bondi M., Brunetti G., Comastri A., Setti G., 2004,
MNRAS, 354, L43
\bibitem{}Brighenti F., Mathews W.\ G., 1997, ApJ, 486, L83
\bibitem{}Brighenti F., Mathews W.\ G., 2006, ApJ, 643, 120
\bibitem{}Campbell-Wilson D., Hunstead R.\ W., 1994, PASA, 11, 33
\bibitem{}Celotti A., Fabian A.\ C., 2004, MNRAS, 353, 523
\bibitem{}Cen R., Ostriker J.\ P., 1999, ApJ, 517, 31
\bibitem{}Chambers K.\ C., Miley G.\ K., van Breugel W.\ J.\ M., 1990,
ApJ, 363, 21
\bibitem{}Chokshi A., 1997, ApJ, 491, 78
\bibitem{}Churazov E., Sunyaev R., Forman W., B{\"o}hringer, H., 2002, MNRAS, 332, 729
\bibitem{}Churazov E., et al., 2005, MNRAS, 363, L91
\bibitem{}Clarke T., Blanton E., Sarazin C., Kassim N., Anderson L., Schmitt H., Gopal-Krishna, Neumann D., 2007, in press in H.\ B{\"o}hringer, P.\ Sch{\"u}cker, G.\ W.\ Pratt, A.\ Finoguenov, eds., Heating vs.\ Cooling in Galaxies and Clusters of Galaxies, Springer Verlag, Berlin (astro-ph/0612595)
\bibitem{}Crenshaw D.\ M., Kraemer S.\ B., 2007, ApJ, in press (astro-ph/0612446)
\bibitem{}Crenshaw D.\ M., Kraemer S.\ B., George I.\ M., 2003, ARA\&A, 41, 117
\bibitem{}Croston J.\ H., Birkinshaw M., Hardcastle M.\ J., Worrall D.\ M., 2004a, MNRAS, 353, 879 
\bibitem{}Croston J.\ H., Hardcastle M.\ J., Birkinshaw M., Worrall D.\ M., 2004b, Nuc.\ Phys.\ B., Proc.\ Supp., 132, 165
\bibitem{}Das S., Chattopadhyay I., Nandi A., Chakrabarti S.\ K., 2001, A\&A, 379, 683
\bibitem{}de Breuck C., van Breugel W., R{\"o}ttgering H.\ J.\ A., 
Miley G. 2004, VizieR On-Line Data Catalog, originally published in 2000,
A\&AS, 143, 303
\bibitem{}de Grandi S., Ettori S., Longhetti M., Molendi S., 2004, A\&A, 419, 7
\bibitem{}De Young D.\ S., 1989, ApJ, 342, L59
\bibitem{}Di Matteo T., Springel V., Hernquist L., 2005, Nature, 433, 604
\bibitem{}Dunn R.\ J.\ H., Fabian A.\ C., 2006, MNRAS, 373, 959
\bibitem{}Eales S.\ A., Rawlings S., Dickinson M., Spinrad H.,
Hill G.\ J., Lacy M., 1993, ApJ, 409, 578
\bibitem{}Fabian A.\ C., et al., 2000, MNRAS, 318, L65 
\bibitem{}Fabian A.\ C., Sanders J.\ S., Crawford C.\ S.,
Ettori S., 2003, MNRAS, 341, 729
\bibitem{}Fang T., et al., 2006, ApJ, 660, L27 
\bibitem{}Ferrara A., Perna R., 2001, MNRAS, 325, 1643
\bibitem{}Ford H., et al., 2004, in D.\ L. Block, K.\ C. Freeman, I.\ Puerari, R.\ Groess, eds., Penetrating Bars through Masks of Cosmic Dust: the Hubble Tuning Fork Strikes a New Note, ASSL 319, Kluwer, Dordrecht, p.\ 459 
\bibitem{}Furlanetto S., Loeb A., 2001, ApJ, 556, 619
\bibitem{}Garrington S.\ T., Conway R.\ G., Leahy J.\ P., 1991, MNRAS, 250, 171
\bibitem{}Garrington S.\ T., Leahy J.\ P., Conway R.\ G., Laing R.\ A., 1988, Nature, 331, 147
\bibitem{}Gergely L.\ A., Biermann P.\ L., 2007, preprint: arXiv:0407.1968
\bibitem{}Gilbert G.\ M., Riley J.\ M., Hardcastle M.\ J., Croston J.\ H., Pooley G.\ G., Alexander P.,
2004, MNRAS, 351, 845
\bibitem{}Gopal-Krishna, 1991, Current Sci., 60, 117
\bibitem{}Gopal-Krishna, Giraud E., Melnick J., della Valle M., 1995, A\&A, 303, 705 
\bibitem{}Gopal-Krishna, Nath B.\ B., 1997, A\&A, 326, 45
\bibitem{}Gopal-Krishna, Swarup, G., 1977, MNRAS, 178, 265
\bibitem{}Gopal-Krishna, Wiita P.\ J., 1996, ApJ, 467, 191
\bibitem{}Gopal-Krishna, Wiita P.\ J., 2000, ApJ, 529, 189 (GKW00)
\bibitem{}Gopal-Krishna, Wiita P.\ J., 2001, ApJ, 560, L115
\bibitem{}Gopal-Krishna, Wiita P.\ J., 2003, in Radio Astronomy at the Fringe, eds.\ J.\ A.\ Zensus, M.\ H.\ Cohen \&
E.\ Ros (ASP Conf. Ser., Vol 300) 293
Anita Publications, New Delhi, p.\ 108 (astro-ph/0409761)
\bibitem{}Gopal-Krishna, Wiita P.\ J., Osterman, M.\ A., 2003, in 
Collin S., Combes F., Sholsmann I., eds, ASP Conf.\ Ser.\ Vol 290, Active Galactic Nuclei: from Central Engine to
Host Galaxy, Astron.\ Soc.\ Pac., San Francisco, p.\ 319 
\bibitem{}Gopal-Krishna, Wiita P.\ J., Barai P., 2004, JKAS, 37, 517
\bibitem{}Gregg M.\ D., Becker R.\ H., de Vries W., 2006, ApJ, 641, 210
\bibitem{}Grimes J.\ A., Rawlings S., Willott C.\ J., 2004, MNRAS, 349, 503
\bibitem{}Hardcastle M.\ J., Alexander P., Pooley G.\ G., Riley J.\ M.,
1997, MNRAS, 288, 859
\bibitem{}Hardcastle M.\ J., Birkinshaw M., Cameron R.\ A., Harris D.\ E., Looney L.\ W., Worrall D.\ M., 2002, ApJ, 581, 
948
\bibitem{}Hardcastle M.\ J., Worrall D.\ M., 2000, MNRAS, 319, 562
\bibitem{}Harvanek M., Hardcastle M.\ J., 1998, ApJS, 119, 25
\bibitem{}Hooda, J.\ S., Wiita, P.\ J., 1998, ApJ, 493, 81
\bibitem{}Hopkins P.\ F., Narayan R., Hernquist L., 2006, ApJ, 643, 641
\bibitem{}Jarvis M.\ J., Wilman R.\ J., R{\"o}ttgering H.\ J.\ A., Binette L., 2003, MNRAS, 338, 263 
\bibitem{}Jenkins C.\ J., Scheuer, P.\ A.\ G., 1976, MNRAS, 174, 327
\bibitem{}Johnson O., Almaini O., Best P.\ N., Dunlop J., 2007,
MNRAS, 376, 151
\bibitem{}Kaiser C.\ R., Alexander P., 1997, MNRAS, 286, 215
\bibitem{}Kaiser C.\ R., Pavlovski G., Pope E.\ C.\ D., Fangohr H.,
2005, MNRAS, 359, 493
\bibitem{}Kapferer W., et al., 2007, A\&A, 466, 813 
\bibitem{}K{\"o}nigl A., Kartje J.\ F., 1994, ApJ, 434, 446
\bibitem{}Krause M., 2005, A\&A, 436, 845
\bibitem{}Kronberg P.\ P., Dufton Q.\ W., Li H., Colgate S.\ A., 2001, ApJ, 560, 178
\bibitem{}Laing R.\ A. 1988, Nature, 331, 149
\bibitem{}Lara L., Cotton W.\ D., Feretti L., Giovannini G., Marcaide J.\ M., M{\'a}rquez I., Venturi T., 2001, A\&A, 370, 409
\bibitem{}Leahy J.\ P., Perley R.\ A., 1991, AJ, 102, 537
\bibitem{}Leahy J.\ P., Williams A.\ G., 1984, MNRAS, 210, 929
\bibitem{}Levine R., Gnedin N.\ Y., 2005, ApJ, 632, 727 (LG)
\bibitem{}Lyutikov M., 2006, MNRAS 373, 73
\bibitem{}Mazzotta P., Edge A.\ C., Markevitch, M., 2003, ApJ, 596, 190
\bibitem{}McCarthy P.\ J., Spinrad H., Djorgovski S., Strauss M.\ A.,
van Breugel W., Liebert J., 1987, ApJ, 319, L39
\bibitem{}McCarthy P.\ J., Spinrad H., van Breugel W., Liebert J.,
Dickinson M., Eisenhardt P., 1990, ApJ, 365, 487
\bibitem{}McCarthy P.\ J., van Breugel W., Kapahi V.\ K., 1991, ApJ, 371, 478
\bibitem{}Meier D.\ L., 2003, NewAR, 47, 667
\bibitem{}Miley G.\ K., et al., 2004, Nat, 427, 47 
\bibitem{}Motl P.\ M., Burns J.\ O., Loken C., Norman M.\ L., Bryan G.,
2004, ApJ, 606, 635
\bibitem{}Narayan R., Yi I., 1994, ApJ, 428, L13
\bibitem{}Norman M.\ L., Burns J.\ O., Sulkanen M.\ E., 1988, Nat, 335, 146
\bibitem{}Omma H., Binney, J., 2004, MRNAS, 350, L13 
\bibitem{}O'Sullivan E., Forbes D.\ A., Ponman T. J., 2001, MNRAS, 328, 461
\bibitem{}Pentericci L., R{\"o}ttgering H.\ J.\ A., Miley G.\ K.,
Carilli C.\ L., McCarthy P., 1997, A\&A, 326, 580
\bibitem{}P{\'e}rez-Torres M.\ A., de Breuck C., van Breugel W., Miley G., 2006, AN, 327, 245
\bibitem{}Peterson J.\ R., Fabian A.\ C., 2006, Physics Reports, 427, 1
\bibitem{}Pounds K.\ A., Reeves J.\ N., King A.\ R., Page K.\ L., O'Brien, P.\ T., Turner M.\ J.\ L., 2003, MNRAS, 345, 705
\bibitem{}Rawlings, S. 2003, NewAR, 47, 397
\bibitem{}Rawlings, S., Jarvis M.\ J., 2004, MNRAS, 355, L9
\bibitem{}Rees M.\ J., 1989, MNRAS, 239, 1P
\bibitem{}Reeves J.\ N., O'Brien P.\ T., Ward M.\ J. 2003, ApJ, 593, L65
\bibitem{}Rowley D.\ R., Thomas P.\ A., Kay S.\ T., 2004, MNRAS, 352, 508
\bibitem{}Sambruna R.\ M., Gambill J.\ K., Maraschi L., Tavecchio F., Cerutti R., Cheung C.\ C., Urry C.\ M., Chartas G., 2004, ApJ, 608, 698
\bibitem{}Saripalli L., Subrahmanyan R., Udaya Shankar N., 2002, ApJ, 565, 256
\bibitem{}Scannapieco E.,  Oh S.\ P., 2004, ApJ, 608, 62 (SO)
\bibitem{}Scannapieco E., Silk J., Bouwens R., 2005 ApJ, 635, L13
\bibitem{}Scheuer P.\ A.\ G. 1974, MNRAS, 166, 513
\bibitem{}Shakura N.\ I., Sunyaev R.\ A., 1973, A\&A, 24, 337
\bibitem{}Silk, J., 2005, MNRAS, 364, 1337
\bibitem{}Soker N.,  Pizzolato F., 2005, ApJ, 622, 847
\bibitem{}Spiegel D.\ S., Paerels F., Scharf C.\ A., 2007, ApJ, 658, 288
\bibitem{}Spinrad H., Dey A., Graham J.\ R., 1995, ApJ, 438, L51
\bibitem{}Statler T.\ S., McNamara B.\ R., 2002, ApJ, 581, 1032
\bibitem{}Sternberg A., Pizzolato F., Soker N., 2007, ApJ, 656, L5
\bibitem{}Storchi-Bergmann T., et al., 2003, ApJ, 598, 956  
\bibitem{}Tormen G., Moscardini L., Yoshida, N., 2004, MNRAS, 350, 1397
\bibitem{}Tremonti C.\ A., Moustakas J., Diamond-Stanic A.\ M., 2007,
ApJ (Lett) in press (arXiv:0706.0527)
\bibitem{}van Oijk R., R{\"o}ttgering H.\ J.\ A., Carilli C.\ L.,
Miley G.\ K., Bremer M.\ N., 1996, A\&A, 313, 25
\bibitem{}van Oijk R., R{\"o}ttgering H.\ J.\ A., 
Miley G.\ K., Hunstead R.\ W., 1997, A\&A, 317, 358
\bibitem{}Venemans B.\ P., et al.\ 2005, A\&A, 431, 793 
\bibitem{}Wiita P.\ J., Norman M.\ L., 1992, ApJ, 385, 478
\bibitem{}Willott C.\ J., Rawlings S., Blundell K.\ M., Lacy M., 1999, MNRAS, 309, 1017
\bibitem{}Wilson A.\ S., Colbert E.\ J.\ M., 1995, ApJ, 438, 62
\end{thebibliography}
\end{document}